\documentclass{ws-p8-50x6-00}

\def\epsi{\mbox{$\varepsilon$}}
\def\epsip{\mbox{$\varepsilon'$}}
\def\reoe{\mbox{${\rm Re}(\epsip/\epsi)$}}

\def\kz{\mbox{$K^0$}}
\def\pz{\mbox{$\pi^0$}}
\def\bkz{\mbox{$\overline {K^0}$}}
\def\kl{\mbox{$K_L$}}
\def\ks{\mbox{$K_S$}}
\def\ppm{\mbox{$\pi^{+} \pi^{-}$}}
\def\ppz{\mbox{$\pi^{0} \pi^{0}$}}
\def\kpp{\mbox{$K \rightarrow \pi \pi$}}
\def\klpp{\mbox{$\kl \rightarrow \pi \pi$}}
\def\klppm{\mbox{$\kl \rightarrow \ppm$}}
\def\ksppm{\mbox{$\ks \rightarrow \ppm$}}
\def\klppz{\mbox{$\kl \rightarrow \ppz$}}
\def\kzppz{\mbox{$\kz \rightarrow \ppz$}}
\def\klppp{\mbox{$\kl \rightarrow \pz \pz \pz$}}

\def\ksppz{\mbox{$\ks \rightarrow \ppz$}}
\def\ketl{\mbox{$\kl \rightarrow \pi e \nu$}}
\def\kmutl{\mbox{$\kl \rightarrow \pi \mu \nu$}}
\def\kets{\mbox{$K_{e3}$}}
\def\kmuts{\mbox{$K_{\mu 3}$}}

\begin{document}

\title{Measurement of Direct CP Violation by NA48}

\author{Giuseppina Anzivino
\footnote{\lowercase{on behalf of the}
NA48 C\lowercase{ollaboration}: C\lowercase{agliari}, 
C\lowercase{ambridge}, CERN, D\lowercase{ubna}, E\lowercase{dinburgh}, 
F\lowercase{errara}, F\lowercase{irenze}, M\lowercase{ainz}, 
O\lowercase{rsay}, P\lowercase{erugia}, P\lowercase{isa},
S\lowercase{aclay}, S\lowercase{iegen}, T\lowercase{orino}, 
W\lowercase{arszawa}, W\lowercase{ien}.}} 
 
\address{Universit\`{a} di Perugia and INFN \\
via A. Pascoli, 06124 Perugia, Italy}

\maketitle

\abstracts{
The NA48 experiment at the CERN SPS aims to search for
direct CP violation in the \kz~system through the measurement 
of \reoe~with high accuracy.
In 1999 the NA48 collaboration has published its first measurement 
based on 1997 data. A new result, based on 1998 and 1999
data, is presented in this article. The result, combined with 
1997 data, $\reoe =  (15.3 \pm 2.6) \times 10^{-4}$, contributes to
the precise determination of the size of direct CP violation.}

\section{Introduction}

In the neutral kaon system the CP eigenstates are linear 
combinations of the strangeness eigenstates, \kz~and \bkz.
If the mass eigenstates, \ks~and \kl, were pure CP 
eigenstates, \kl~would decay only into CP=--1 and \ks~only
into CP=+1 final states. The two decay modes in \ppm~and \ppz~
have CP=+1, so \ks~are allowed to decay into pion pairs, but not \kl.
Then, in the neutral kaon system CP violation manifests in the 
observation of the CP-forbidden \klpp~decays.
First evidence of CP violation was observed in 1964 by 
Christenson, Cronin, Fitch and Turlay\cite{CP64}.

In the Standard Model, CP violation is related to the existence
of three generations of quarks and to a complex phase in the CKM
matrix. Two different mechanisms contribute: {\it indirect}  
CP violation, due to the mixing of the \kz~and \bkz~states, 
represented by the parameter \epsi, and {\it direct} CP violation, 
due to the decay process itself, through the interference
of final states with different isospins, and represented 
by the parameter \epsi'.

The parameters \epsi~and \epsi' are related to the amplitude 
ratios

$\eta_{+-} = \frac{A(\klppm)}{A(\ksppm)}$ $\simeq \epsi + \epsi'$
\hspace{0.5cm}
$\eta_{00} = \frac{A(\klppz)}{A(\ksppz)}$ $\simeq \epsi - 2\epsi'$ 

which represent the strength of the CP violating amplitudes
with respect to the CP conserving ones, in each mode.
\epsi' is small compared to \epsi~and it is convenient 
to measure $\reoe$. 
The experimental observable, the double ratio $\rm R$,
is related to the decay widths and to \reoe~through

\begin{center}
$\mathrm{R} = {\frac{\Gamma (\klppz)}{\Gamma (\ksppz)}/
      \frac{\Gamma (\klppm)}{\Gamma (\ksppm)}} 
      \simeq  1 - 6\ {\reoe }$
\end{center}
 
The first generation of experiments, NA31 at CERN and E731 at
Fermilab, gave inconclusive results: while NA31\cite{NA31}
reported evidence for direct CP violation, indicating a $3.5 \sigma$
effect,  the result of E731\cite{E731} was compatible with no effect. 
Two new experiments were setup to clarify the situation. 

In 1999 first results were reported and both experiments 
confirmed the existence of direct CP violation: 
KTeV at Fermilab  measured 
$\reoe = ((28.0 \pm 4.1)\times 10^{-4})$\cite{KTeV} 
and NA48 at CERN $\reoe = ((18.5 \pm 7.3)\times 10^{-4})$\cite{NA48}. 
Before June 2001, the four most precise results
\cite{NA31,E731,KTeV,ivan}  
gave a world average of $(19.2\pm2.5)\times 10^{-4}$,
with a  $\chi^2$/ndf = 10.4/3, that indicates poor constincency
between the data.
In order to establish the size of direct CP violation, new 
results form the two experiment, with substantially smaller
uncertainties, were expected. Current theoretical predictions
are in the range ($1 \div 30 \times 10^{-4}$)\cite{teo}.

This paper describes a new result from NA48, based on the analysis 
of data collected during the 1998 and 1999 runs; the 
statistics, is seven times larger ($\sim 3.3\times10^6\, \klppz$)
than that used for the published 1997 result\cite{NA48}.

\section{The NA48 method}

NA48 aims to measure \reoe~with an accuracy of~ $2\times10^{-4}$
using the double ratio technique and nearly collinear simultaneous
\ks~and \kl~beams. The required statistical error is reached 
by collecting several millions of \klppz~(the statistically
limiting decay mode); data are collected using fast and efficient 
data acquisition system, triggers with high rejection power and
high capacity data storage systems.

The basic principle of the experiment is to make the systematic
biases symmetric between either the \ppm~and \ppz~decays  
or the \kl~and \ks~beams. In this way most of the systematic 
effects cancel to first order and only the differences between 
two components need to be considered in detail in the analysis.

In order to fully exploit the double ratio technique, the four 
decay modes are collected
{\it simultaneously}; this cancels out effects due to beam fluxes,
inefficiencies in \ks, \kl~identification, trigger and event
selection, as well as dead time and accidental losses.
In order to minimise acceptance corrections, the four decay modes
are collected 
from a {\it common decay region} and with two
{\it quasi collinear beams}, so that the decay products illuminate
the detector in a similar way. However, due to the very different 
lifetimes of the \ks~and \kl~particles,
the vertex distributions vary a lot along the beam direction.
In order to obtain almost identical decay distributions 
for \kl~and \ks, \kl~decays are {\it weighted}  with 
a function of the proper time.
The backgound, that cannot be reduced by cancellation, is
suppressed using {\it high resolution detectors}.
The remaining \ks-\kl~difference in energy spectra are reduced
by performing the analysis in {\it energy bins} (20 bins, between
70 and 170 GeV).

\section{Experimental set-up}

The NA48 beam line\cite{beams} and detector\cite{det} are
designed to fulfill these requirements.
A primary 450 GeV proton beam ($\sim 1.5 \times 10^{12}\, ppp$)
at the SPS accelerator produces the \kl~beam on a beryllium
target. The non-interacting protons are sent to a bent 
silicon crystal\cite{bent}; only a small fraction of 
protons satisfy the conditions for channelling and 
an attenuated proton beam ($\sim 3 \times 10^{7}\, ppp$)
impinges on the \ks~target. In the way to the target, the
protons pass through a tagging station, that tag protons 
that are going to produce the \ks~beam.

The decays in \ppm~are reconstructed using a magnetic
spectrometer composed of four large drift chambers and
a dipole magnet. The momentum resolution is  
$\sigma (p)/p \simeq 0.5 \% \oplus 0.009 p[GeV/c]\% $
($\sim 1 \%$ for 100 GeV/c track momentum).
The time of~ \ppm~events is measured by a scintillator 
hodoscope ($\sigma_{t} \approx 200\, ps$). 
The trigger for \ppm~decays consists of a fast pretrigger
and a processor farm\cite{mbox} that computes the decay vertex
position and the invariant mass from the drift chamber signals.
This trigger has an inefficiency of $\sim 2.2 \%$ and a dead time
of $\sim 1.1 \%$. 

The decays in \ppz~are reconstructed from the informations on
energy, position and time of the four clusters given by the
Liquid Krypton electromagnetic calorimeter. The energy resolution is
$\sigma (E)/E \simeq 3.2\%/\sqrt{E} \oplus 
0.09/E \oplus 0.42\%$ (E in GeV, better than 1\% for  25 GeV photons).
The time resolution is $\sim 250\, ps$.
The neutral trigger\cite{nut} uses the calorimeter informations 
and look-up tables to make a fast decision. The inefficiency
is $\sim 0.1 \%$ with almost no dead time.

The reconstructed decay products have to be assigned to a parent
particle (either \kl~or \ks); the identification is done using
the tagging station. The time of flight between the detector
and the tagger is measured: events with a time coincidence 
inside a $\pm 2 ns$  window are assigned to \ks~parent particle,
events outside this window are identified  as \kl.

\section{Data analysis}

The data collected during the 1998 run amount to 1.1 
millions of \klppz~in 135 days of data taking. In 1999,
an upgrade of the trigger and event builder PC farm,
as well as a better stability of the detectors and
electronics, resulted in high data taking efficiency
and the experiment collected, in 128 days, 2.2 millions
of \klppz. 

\subsection{Event selection}

The analysis is based on the principle of
minimising the corrections to be applied to 
the double ratio of the decay counts. 
The key points of the analysis
can be summarized as follows:
\begin{itemize}
\item The four decay modes are counted in the same kaon
energy interval ($70\,GeV<E_K<170\,GeV$), and decay volume
($0<\tau_S<3.5$ in units of \ks~lifetime). An anti-counter 
(AKS), placed
in the \ks~beam, determines the beginning of the \ks~decay
region. 
\item Dead time in the trigger or read out is applied to
all four modes, in order to equalise intensity conditions and
to preserve the principle of symmetrization.
\item In \ppm~decays \ks~and \kl~can also be identified by 
extrapolating
the vertical decay vertex position. To keep the symmetry
principle, the tagging is used for \ks~-\kl~identification
both for \ppz~and \ppm~decays; in this way
the measurement is sensitive only to differences
between neutral and charged mistagging probabilities.
\item To cancel the contribution from the different 
\kl~and \ks~lifetimes to 
the acceptance, each \kl~candidate is weighted 
with a function of
the proper time to match the decay vertex 
distribution of \ks~(Fig. 1).
\end{itemize}

Events are counted, in each sample, and corrections are applied
in twenty bins of kaon energy, each 5\,GeV wide, 
in order to reduce the influence of residual difference in
\ks/\kl~spectra. The result is obtained by averaging the 
twenty double ratios.

\subsection{\ks-\kl~misidentification}

\begin{figure}
\vspace{-0.5cm}
\begin{minipage}{0.49\textwidth}
 \includegraphics[width=\textwidth]{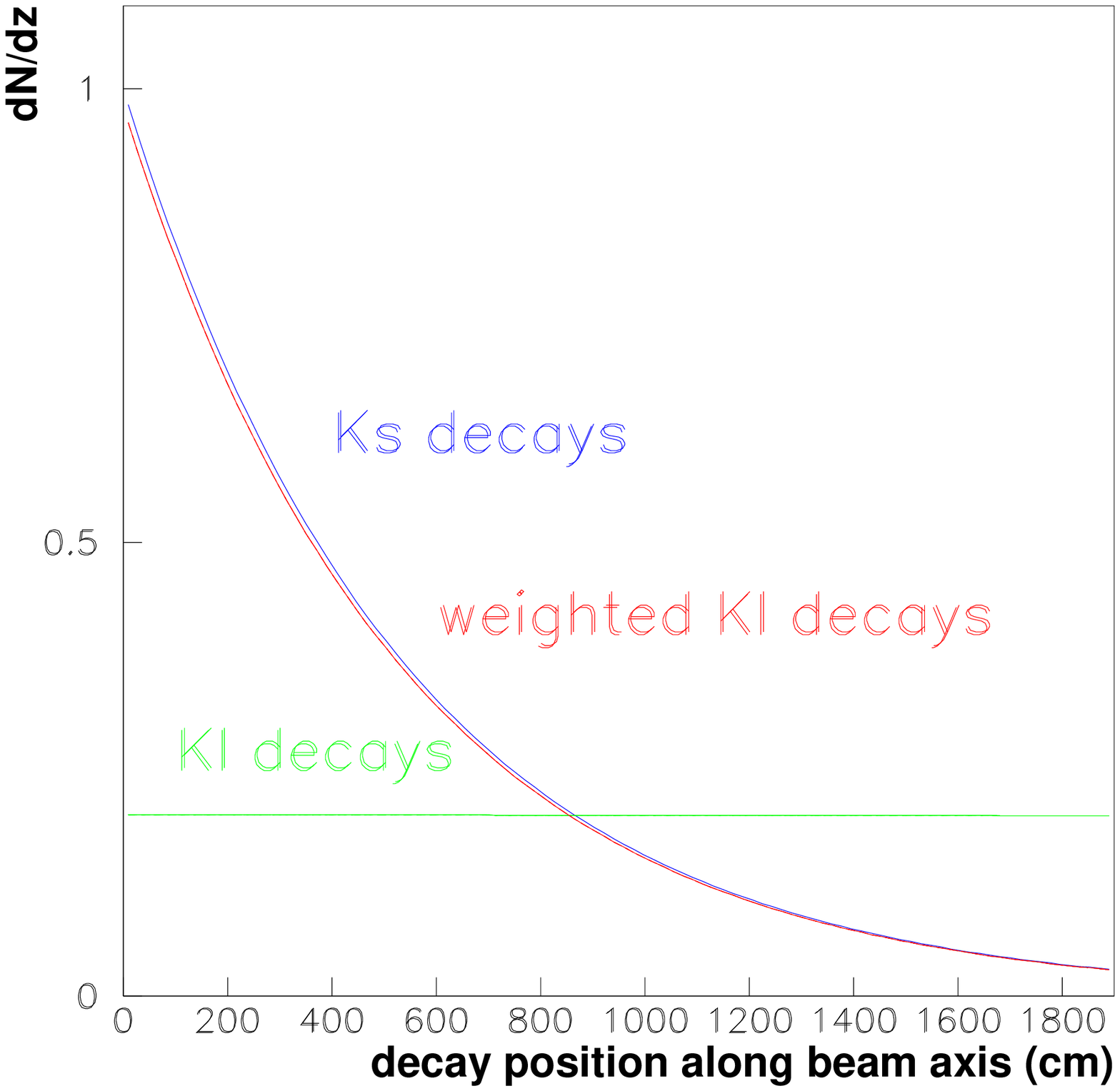}
  \end{minipage} 
\begin{minipage}{0.49\textwidth}
 \vspace{0.6cm}
 \hspace{0.3cm}
 \includegraphics[width=\textwidth]{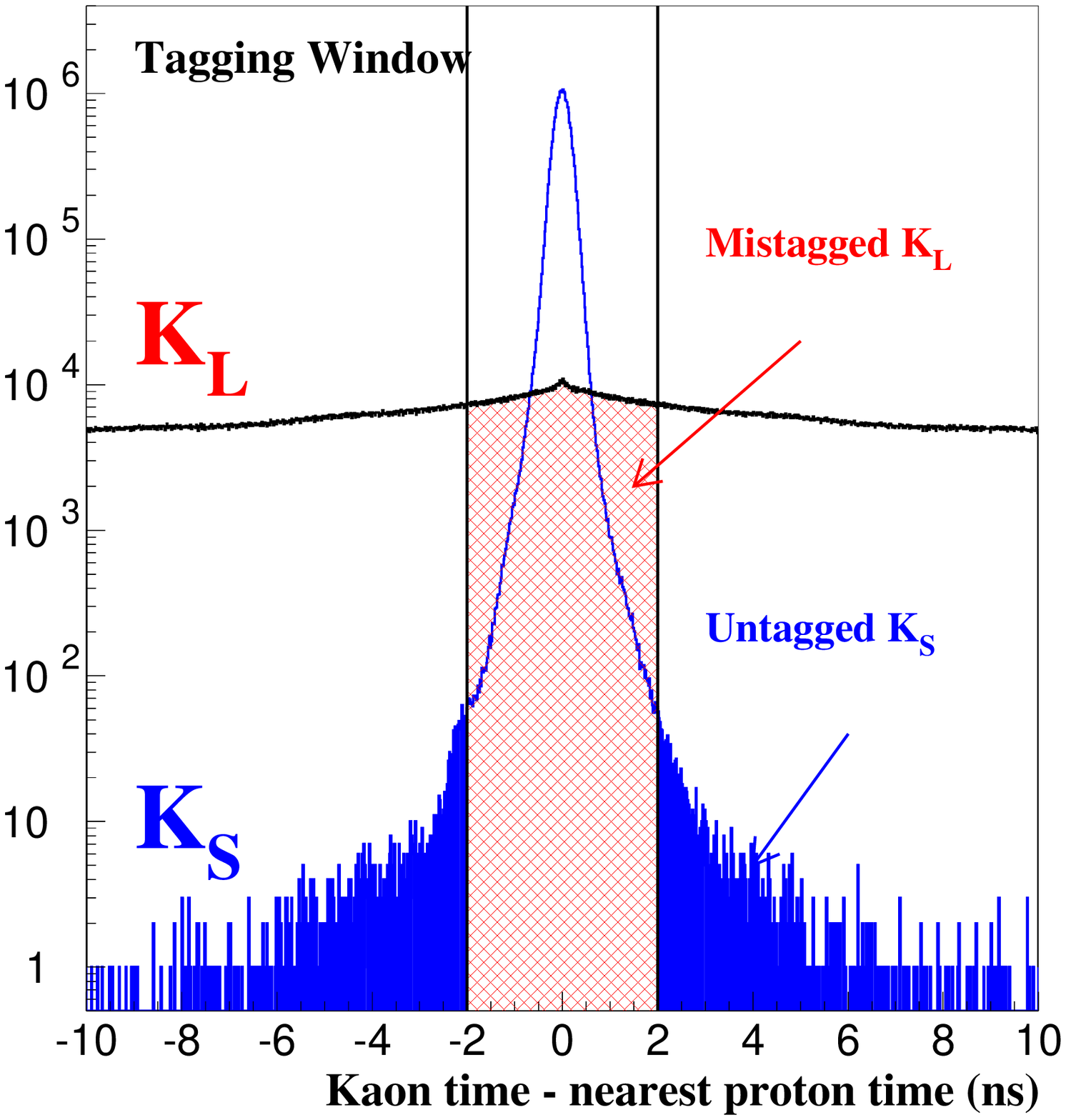}
  \end{minipage}
\caption{left: Reconstructed lifetime distribution for \klpp~before
and after weighting;
right: The time difference between a \kpp~candidate
and the nearest proton time in the tagger.
The \ks~and \kl~contributions are
identified by vertical vertex separation. \label{fig:tag}}
\end{figure}

In the tagging procedure, there are two possible sources of 
misidentification: \ks~can be identified as \kl~($\alpha_{SL}$), 
due to tagging inefficiencies, and \kl~can be identified as 
\ks~($\alpha_{LS}$), due to accidental
coincidences between protons in the tagger and \kl~decays.
The double ratio is sensitive only to the differences 
between the mistagging probabilities,
$\Delta\alpha_{SL} =\alpha_{SL}^{00} - \alpha_{SL}^{+-}$ and
$\Delta\alpha_{LS} =\alpha_{LS}^{00} - \alpha_{LS}^{+-}$.
The tagging inefficiency measured in \ppm~decays is
$\alpha_{SL}^{+-}=(1.63 \pm 0.03)\times10^{-4}$; the differential
effect is evaluated using 2\pz~and 3\pz~events with one photon
conversion: $\Delta\alpha_{SL} = (0 \pm 0.5) \times 10^{-4}$.
The accidental tagging probability, measured
with \ppm~decays, is $\alpha_{LS}^{+-}=(10.649\pm0.008)$;
the value of $\Delta\alpha_{LS}$ is estimated using sidebands
away from the coincidence peak in the time difference distribution:
$\Delta\alpha_{LS} = (4.3 \pm 1.8) \times 10^{-4}$ (Fig. \ref{fig:tag}).

\subsection{Charged and Neutral Background}
In both decay modes, \ks~has no significant background.
The main source of background in \ppm~decays are semileptonic
3-body decays \ketl~and \kmutl. The \kets~is rejected by 
measuring the energy of the electrons in the calorimeter and 
their momentum in the spectrometer and then by requiring
$E/p<0.8$. For the \kmuts~case, no associated hit in the 
muon veto is required. In addition a cut on the kaon 
invariant mass is applied. The remaining background is subtracted 
using two control regions in the invariant mass-rescaled transverse
momentum squared plane (${p_{T}^{'}}^2$). 
The background under the signal region
is estimated using cleanly identified \kets~and \kmuts~events.

\begin{figure}
\vspace{-0.5cm}
\begin{minipage}{0.49\textwidth}
 \includegraphics[width=\textwidth]{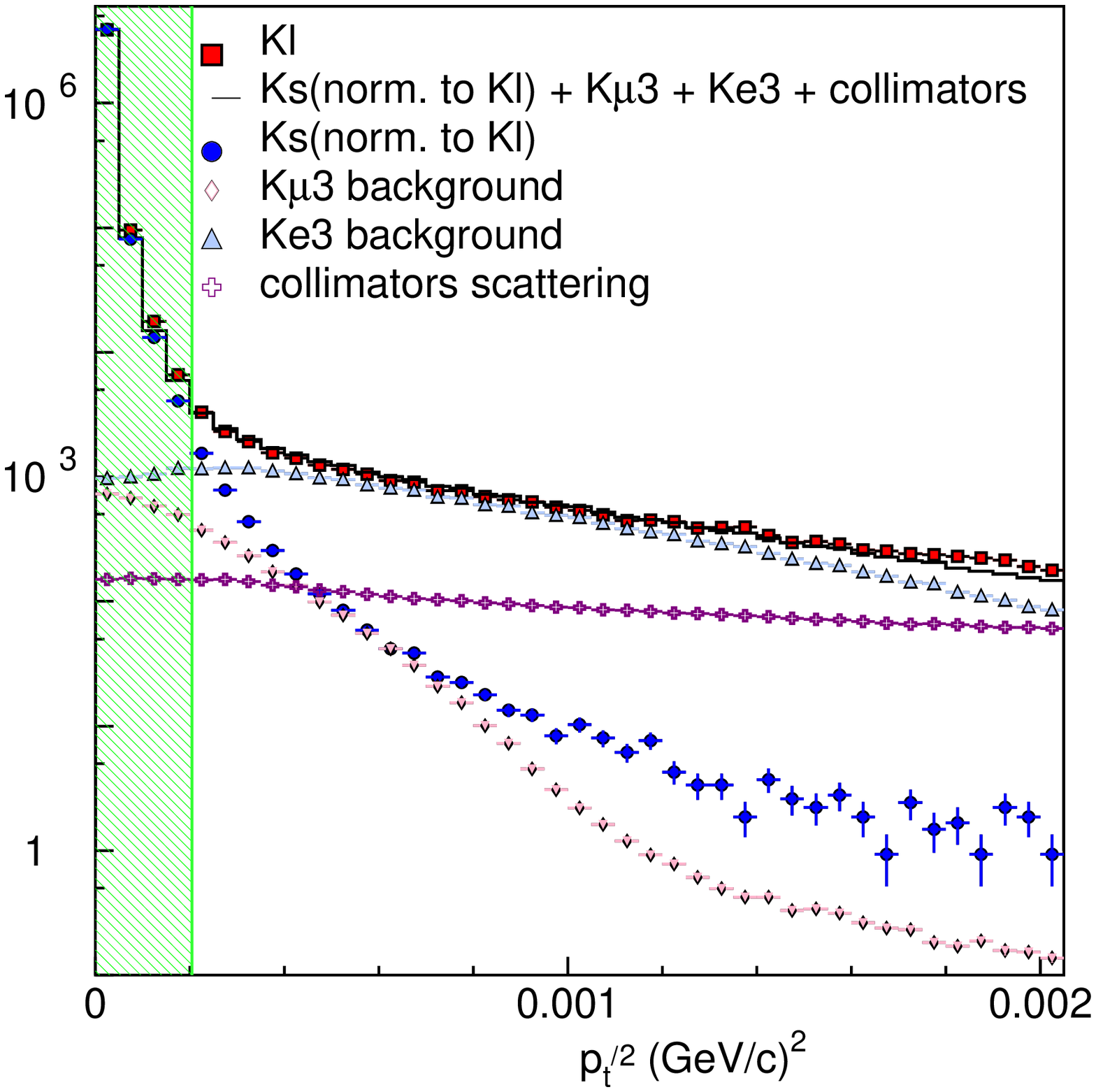}
  \end{minipage} 
\begin{minipage}{0.49\textwidth}
  \hspace{0.3cm}
 \includegraphics[width=\textwidth]{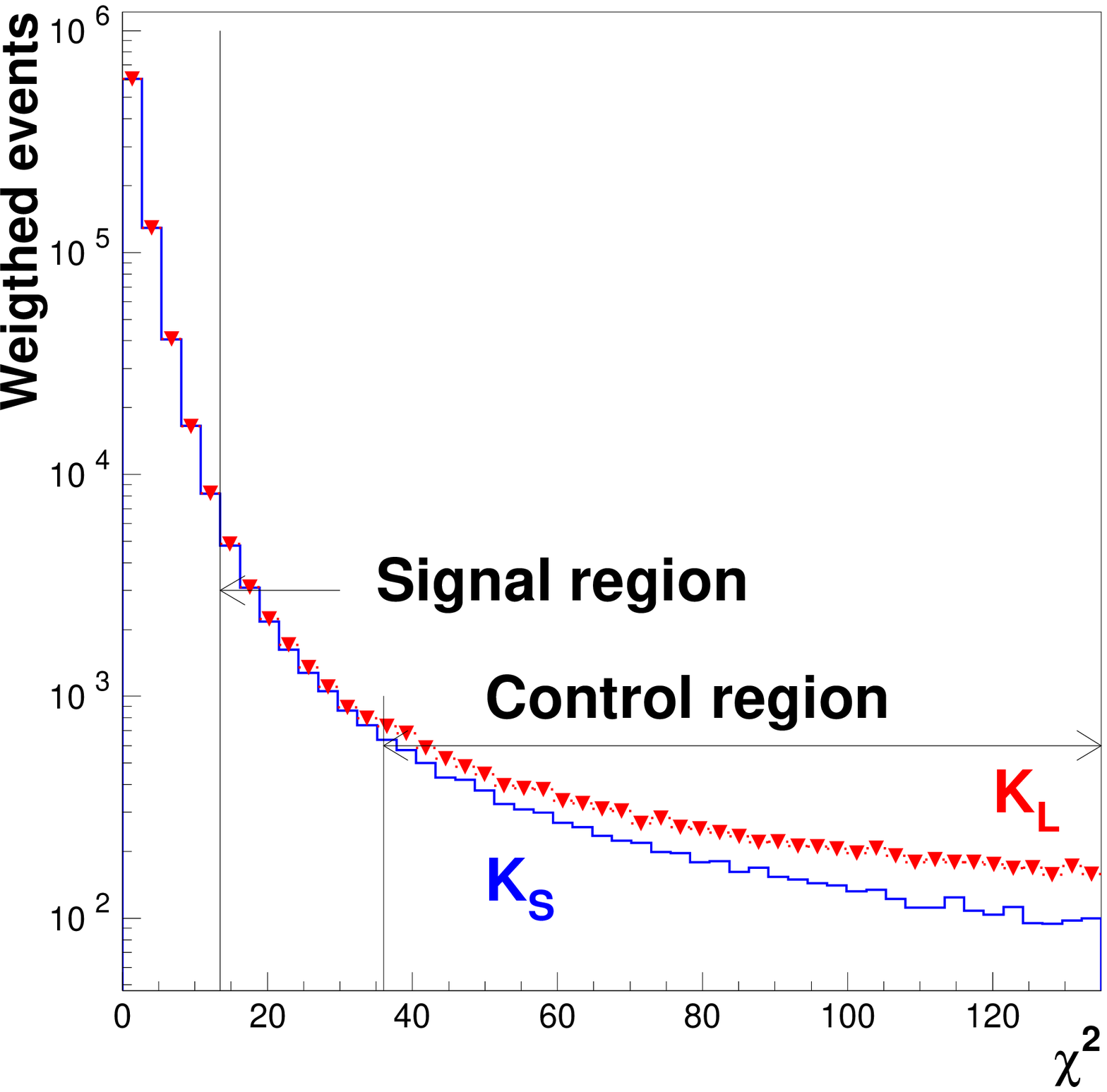}
  \end{minipage}
\caption{left: Distribution of charged signal and background
components in the ${p_{T}^{'}}^2$; 
right: Distribution of neutral 
signal and background excess in \kl~in the $\chi^2$ variable. 
\label{fig:bkg}}
\end{figure}

In \ppz~decays, the dominant background is \klppp.
This background is rejected by requiring no extra showers
in time. Furthermore an invariant mass cut in the space 
of the two \pz~invariant mass is applied, using a 
$\chi^2$-like variable. The residual  backgroung under the signal 
is estimated from the control region in the $\chi^2$ distribution
(Fig. \ref{fig:bkg}).

\subsection{Other systematics}

Other sources of systematics have been carefully studied. 
The {\it accidental activity} effects are evaluated by overlaying
random events to real good events and by studying the beam
correlation at small and large time scales.
In the \ppz~decays the distance scale is directly
related to the {\it energy scale}. The latter is constrained by
reconstructing the well known position of a detector: in our case 
it is the AKS position, which determines the 
beginning of the decay region.
The {\it acceptance correction} is reduced to a very small level
due to the symmetric detector illumination, consequence
of the \kl~weighting procedure. The small residual effect,
due to the small angle between the two beams, is estimated
using Monte Carlo. Table \ref{tab:sys} shows a list of all 
corrections and systematic uncertainties applied to
the raw double ratio; the effect on \reoe~is obtained 
dividing the numbers by a factor of 6.

\begin{table}
\begin{center}
\begin{tabular}{|c|rl|} \hline
Source  & \multicolumn{2}{c|}{$\Delta(R)$ (in $10^{-4}$ units)} \\ \hline
$\pi^+\pi^-$ background &  {$16.9$} & {$\pm3.0$} \\
$\pi^0\pi^0$ background &  {$-5.9$} & {$\pm2.0$} \\
beam scattering background & {$-9.6$} & {$\pm2.0$} \\
Tagging inefficiency       &   & {$\pm3.0$} \\
Accidental tagging         & {$8.3$} & {$\pm3.4$} (part. stat)\\
$\pi^+\pi^-$ scale  & {$2.0$} & {$\pm2.8$} \\
$\pi^0\pi^0$ scale  &  & {$\pm5.8$} \\
AKS inefficiency    & {$1.1$} & {$\pm0.4$} \\
Acceptance correction & {$26.7$} & {$\pm4.1$} (MC stat) \\
                      &                & {$\pm4.0$} (syst) \\
$\pi^+\pi^-$ trigger  & {$-3.6$} & {$\pm5.2$} (stat)\\
Accidental event losses &              & {$\pm4.4$} (part. stat)\\
\hline
Total & {$35.9$} & {$\pm$12.6} \\
\hline
\end{tabular}
\caption{Corrections and systematic uncertainties on R.
\label{tab:sys}}
\end{center}
\end{table}

\section{The result}

The result, obtained from the data collected in 
years 1998 and 1999 (statistics shown in Table \ref{tab:stat}), is

\begin{center}
$\mathrm{R} = 0.99098 \pm 0.00101_{stat} \pm 0.00126_{syst}$
\end{center}

which corresponds to

\begin{center} 
$\reoe = (15.0 \pm 2.7) \times 10^{-4}$
\end{center}

Combining this result with the published result 
(1997 data)\cite{NA48}, we obtain

\begin{center}
$\reoe = (15.3 \pm 2.6) \times 10^{-4}$
\end{center}

\begin{figure}[h]
\begin{minipage}{0.49\textwidth}
 \includegraphics[width=\textwidth]{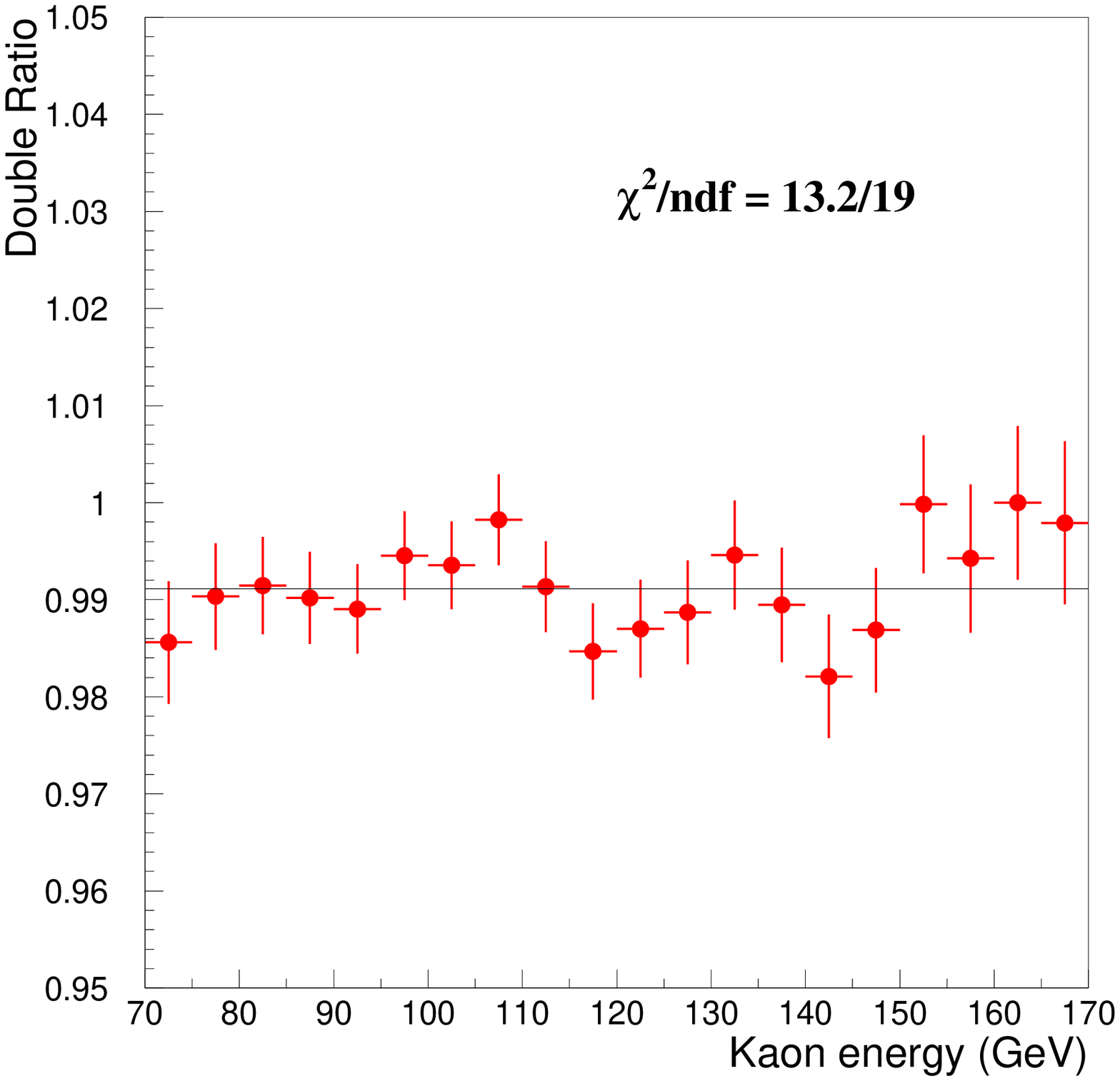}
  \end{minipage} 
\begin{minipage}{0.49\textwidth}
  \hspace{0.3cm}
 \includegraphics[width=\textwidth]{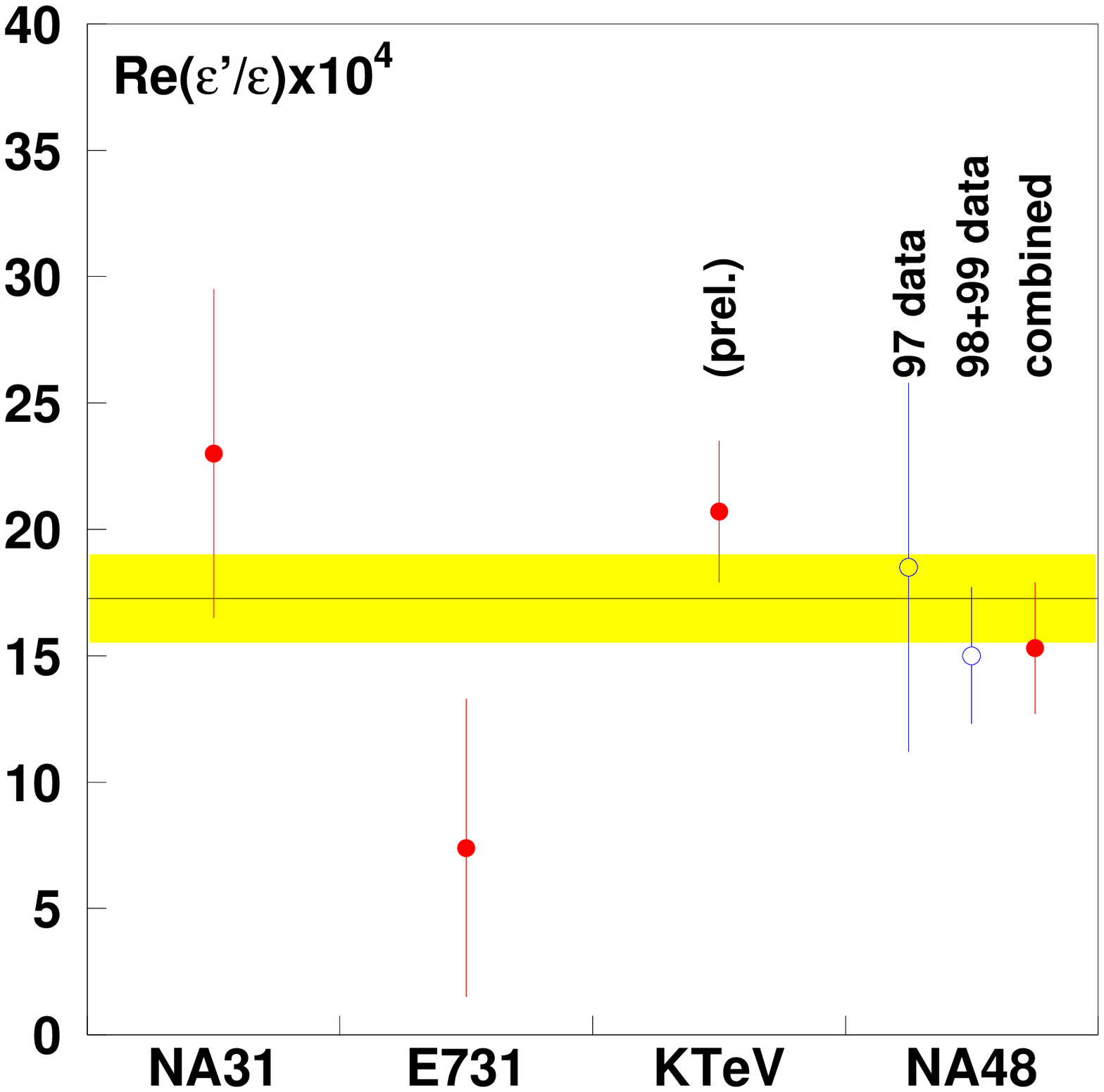}
  \end{minipage}
\caption{left: Stability of the double ratio with kaon energy;
right: Experimental results comparison; full points are used in 
the average. \label{fig:checks}}
\end{figure}

The stability of the corrected double ratio with applied
cuts and changes in beam and detector conditions has been
extensively checked.
In Fig. \ref{fig:checks} the stability with kaon energy is shown.

\begin{table}[h]
\begin{center}
\begin{tabular}{|l|r|l|r|}
\hline
$\klppz$ & 3.29 $\times10^6$ &
$\ksppz$ & 5.21 $\times10^6$ \\ \hline

$\klppm$ & 14.45 $\times10^6$ &
$\ksppm$ & 22.22 $\times10^6$ \\
\hline
\end{tabular}
\caption{Number of selected events after accounting for mistagging.
\label{tab:stat}}
\end{center}
\end{table}

\section{Conclusion and outlook}

The NA48 combined result is 5.9 standard deviations away 
from zero. This confirms the existence of direct CP violation,
with a positive value of \epsi'. 

Recently the KTeV collaboration has anounced a new 
result\cite{KTeV-pisa} based on data from 
the 1997 run and a revised analysis of their published 
result\cite{KTeV} (part of the 1996 statistics):
$\reoe = (20.7 \pm 2.8) \times 10^{-4}$.

Taking into account the results from previous generation
expriments \cite{NA31,E731}, the KTeV updated 
result\cite{KTeV-pisa} and the NA48 combined result, 
presented in this paper, we obtain a world average 
(see Fig. 3) of 

\begin{center}
$\reoe = (17.2 \pm 1.8) \times 10^{-4}$
\end{center}

with $\chi^2$/ndf = 5.5/3 (14 $\%$ probability).

This definitely establishes direct CP violation.

In the year 2000, NA48 collected data without the spectrometer,
due to an implosion of the beam pipe that damaged all 
the four drift chambers; these data were used for systematic
checks for the neutral decays. 

The year 2001 is the last year of data taking and we expect
to collect $\sim 1.5$  millions of \klppz. We will complete the
statistics and, with the new spectrometer
and different beam conditions (spill), we will check the
result under different conditions.

\end{document}